\documentclass[aps,prc,twocolumn]{revtex4}
\usepackage{graphicx}

\begin{document}
\title{Alternative model of propagation of spikes along neurons}

\author{Marat~M.~Rvachev}
\email{maratr@mit.edu, rvachev@alum.mit.edu}
\affiliation{Physics Department, Massachusetts Institute of Technology, Cambridge, Massachusetts 02139, USA}

\date{\today}

\begin{abstract}
Here a viable and never before investigated mechanism of propagation of spikes along neurons
is proposed. In the following,
velocity of propagation of small-amplitude pressure waves through the
cytoplasmic interior of myelinated and unmyelinated axons of different
diameters is theoretically estimated, and is found to generally agree with
the action potential (AP) conduction velocities. This remarkable coincidence
allows to surmise a model in which AP spread along axon is
propelled not by straggling ionic currents as in the widely accepted local
circuit theory, but by mechanoactivation of the membrane ion channels by a
traveling pressure pulse. Hydraulic pulses propagating in the viscous
axoplasm are calculated to decay over $\sim$ 1 mm distances, and it is further
hypothesized that it is the role of influxing during the AP
Ca$^{2+}$ ions to activate the membrane skeletal protein network 
for a brief radial contraction amplifying the
pressure pulse and preventing its decay. 

The model correctly predicts that the AP conduction velocity
should vary as the one-half power of axon diameter for large unmyelinated
axons, and as the first power of the diameter for myelinated axons,
provided that specific mechanical properties of axons are independent
from diameter; that myelinization increases the conduction velocity;  
that the conduction velocity increases with the temperature.  Unlike the
local circuit theory, the model is able to qualitatively explain observed
increase in the AP duration in axons of smaller diameters.
Predictions of absolute AP conduction velocities are limited by the knowledge
of relevant to propagation of pressure waves mechanical properties of axons, 
still, the velocities are predicted well for
myelinated axons, while an agreement for unmyelinated axons requires
3 orders of magnitude higher resistance of axons to deformation increasing
their diameter, compared to values deduced from published data on
membrane area expansion moduli. Experimental test of the model is needed.
\end{abstract}

\maketitle

Depolarization of interior of an excitable cell to a critical level leads
to spiking behavior of voltage across the cell membrane, or action
potential (AP). The spike has been shown to arise from rapid changes in
membrane ion-specific permeability, allowing flow of transmembrane ionic
currents powered by electrochemical gradients \cite{hill77}. Membrane permeability
to ions was found to be regulated by voltage across membrane, and
a self-regenerating scheme of time course of the AP was developed \cite{hodg52}.
The rapid changes in membrane permeability were later shown to arise from
brief activation of voltage-gated ion-selective protein channels embedded in
the membrane \cite{shep94}.  These findings are supported by more than half a century of
experiments and are a basis of modern research in neuroscience \cite{shep94}.

Phenomenon of propagation of the AP through excitable cells 
(as opposed to development of the AP at a given cell point) has been explained
by the local circuit theory, which postulates that spread of ionic
currents powered by AP-associated voltage spike across membrane
depolarizes adjoining unexcited membrane and brings it to the critical level for
excitation \cite{hodg52}. Prediction of velocity of AP conduction is an important test of
this model, and can be found from solution of a second-order
partial differential equation for time-dependent potential along the cell length,
with parameters such as electrical permeability of cytoplasm and membrane, and membrane capacitance \cite{hodg52}. A simplified form of
this equation was numerically solved in \cite{hodg52} for a squid giant axon model,
with membrane permeability to Na$^+$ and K$^+$
parametrized as a function of voltage history across membrane, as
deduced from voltage-clamp data. The conduction velocity was found to be
18.8 m/s for a 476 $\mu$m diameter axon at 18.5 $^\circ$C with 21.2 m/s experimental
value.

The local circuit theory predicts that conduction velocity in unmyelinated
nerve fibers should vary as the one-half power of axon diameter if
axoplasmic resistivity and membrane electrical properties per unit area are constant, 
with experimental data fitting powers between 0.57 and 1 depending on
the fiber class \cite{hill77}. For myelinated fibers, the local circuit theory predicts 
that the velocity should vary as the first power of diameter, again with
provisions of constant specific electrical properties, and with additional constraint of
proportional scaling of
axon diameter, external myelin diameter, distance between the nodes of
Ranvier and membrane area at the nodes \cite{rush51}. The first power relationship is
well established for large myelinated fibers \cite{hill77}.

Experiments show that in smaller diameter
axons of a related class, duration of the AP spike becomes longer, a fact completely unexplained by the local
circuit theory \cite{hill77}. Conduction velocity increases with the temperature, e.g. for
cat vagus myelinated fibers Q$_{10}$ (ratio of the velocity at one
temperature to the velocity at a temperature 10 $^\circ$C lower) is 4.8 at 18 $^\circ$C, 2.5
at 28 $^\circ$C and 1.6 at 37 $^\circ$C \cite{paint65}, and for 
desheathed rabbit vagus unmyelinated
fibers Q$_{10}$ is 3.5 at 10 $^\circ$C, 2.1 at 20 $^\circ$C and 1.7 at 
30 $^\circ$C \cite{hill77}. These changes are
explained by the local circuit theory only qualitatively, through
increased rate of conformational changes in the membrane
channel proteins at higher temperatures \cite{hill77}.

Overall, except for the numerical calculation of the conduction velocity for
the squid giant axon model \cite{hodg52}, agreement between local circuit theory
predictions and measurements is close, but rather loose. Agreement in dependence
of the velocity on axon diameter for unmyelinated fibers requires variation in
specific electrical properties of axoplasm or membrane with diameter, while for
similar agreement for myelinated fibers these are assumed to be constant
with (experimentally not strictly satisfied) assumptions of geometrical scalings \cite{hill77}. 
Temperature affects conduction
velocity similarly in myelinated and unmyelinated fibers, while if due only
to changes in activation rate of membrane channels, simple reasoning
suggests that myelinated fibers would be affected much less, since their
conduction velocity would change at the Ranvier nodes only. As already
mentioned, longer duration of the AP in fibers of smaller diameters is not
explained.

The purpose of this paper is to show that by making two
assumptions, namely, that a hydraulic pulse propagating through
axoplasm along axon length is able to mechanoactivate enough membrane ion channels for 
the influxing currents to
depolarize membrane to the excitation level, and that at least one of
influxing during the AP ion species, presumably Ca$^{2+}$, activates the 
membrane skeletal protein network for a brief contraction timed to amplify
the propagating hydraulic pulse, one obtains a competing with the local circuit
theory model of AP propagation, and that this model
agrees reasonably well with experimental data on the AP conduction
velocities. 

The exact mechanisms of how a traveling hydraulic pulse may be amplified and
how it might elicit membrane depolarization are uncertain and most
appealing to author processes are discussed here. However, exact
nature of these processes does not change basic model features and
conclusions drawn, as long as these processes do amplify the hydraulic
pulse and do excite the membrane upon arrival of the pulse. Just as in the
local circuit theory, it is assumed that depolarization of a membrane segment to the
critical for excitation level causes
membrane voltage to spike due to activation of membrane voltage-gated ion channels. 
The distinction from the local
circuit theory is that this initial depolarization is brought about by a
propagating hydraulic pulse and not by the straggling ionic currents, and thus
velocity of AP conduction is equal to velocity of 
propagation of the hydraulic pulse. 
It should be noted that correct prediction of AP shape by the Hodgkin-Huxley equations \cite{hodg52}
does not immediately refute the present model, since local development of AP is the same in both mechanisms.

Although most of discussion here is concerned
with the AP propagation in neurons, the model
is easily generalized to non-neuronal excitable cells.
\section{Model features and necessary assumptions}
Hydraulic pulses do propagate through viscous liquids enclosed in
flexible tubes, a fact illustrated by blood pulse propagating through blood vessels. 
In addition to normal force on the tube wall exerted by
the pulse, viscosity of the liquid introduces shear stress in the direction
parallel to the axis of the tube \cite{morg54}. These forces acting on an extensible
axolemma should lead to a membrane stretch, which, according to the first
assumption, alone or in combination with bending and compression of the
membrane by the propagating pulse, should open enough ion channels to
bring membrane voltage to the excitation level. Evidence is now
mounting that many unrelated types of ion channels are (for unclear reasons) mechanosensitive
\cite{gu01}, but still it is not here possible to quantitatively
substantiate this assumption, partly because magnitude of the membrane
deformations depends on an unknown pulse amplitude.  Since many
mechanosensory ion channels are activated by protein links anchoring channel
proteins to extracellular structures and to the cytoskeleton \cite{gill01}, it is
also reasonable to propose that perturbation of the cell interior cytoskeleton by the
propagating hydraulic pulse could be directly mechanically coupled by
the protein links to the channel proteins, for a fast and sensitive
activation.

As will be shown below, pressure waves in axoplasm decrease $e$-fold in
amplitude over $\sim$ 1 mm distances, so that their sustained propagation in long
axons would require amplification. It appears that forceful circumferential 
(and consequently radial) contraction of the membrane
skeletal protein network, triggered by influxing during the AP Ca$^{2+}$ ions, is 
the most plausible mechanism
of such amplification.  Indeed, virtually every excitable cell has
voltage-gated Ca$^{2+}$ channels in its membrane and, in fact, calcium is the
basis of the AP in muscles of many invertebrates, in smooth muscles and
in many gland cells of vertebrates \cite{shep94}. At the same time, increase in free
intracellular calcium is often associated with initiation of
motion, from motility in freely moving cells and muscle contraction to
synaptic vesicle release in synapses \cite{shep94}. Influxing during the AP calcium
would instantly manyfold increase its concentration in the membrane skeletal protein
network near the inner membrane surface and that, coupled with calcium ability to quickly
induce conformational changes in proteins, such as upon its binding
to actin filaments in actin-myosin muscle complex, could provide for a
fast contractile response necessary to amplify a quickly propagating pressure
pulse. The well-known sodium impulse in this scheme could simply be the
means to rapidly change transmembrane voltage by the sodium current and
thus quickly open relatively sparse voltage-gated calcium channels.

The purpose of the voltage spike in this model is to activate
voltage-gated calcium channels, but clearly the spike will create
axoplasmic ionic currents that, just as in
the local circuit theory, will depolarize
adjacent unexcited membrane. This depolarization, then, will decrease
number of mechanoactivated by the hydraulic pulse ion channels  
needed to bring the membrane to the
excitation voltage; but if the straggling currents are able to excite
adjacent membrane before arrival of the pressure pulse, AP will obviously
spread according to the local circuit theory mechanism. Thus, it
appears that the two mechanisms of propagation can be viewed as competing,
and it is conceivable that different cells, depending on their particular
electrical and mechanical properties, are able to realize one or the other
mechanism.

Another argument in favor of pressure pulse amplification by structures
adjacent to the membrane inner surface is the existence of (currently functionally unexplained) dense
``undercoating'' just beneath axon membrane in the axon hillock and in the nodes of
Ranvier \cite{pala77}.  Initiation of a hydraulic pulse in the axon hillock, as well as
restoration of its amplitude in the nodes of Ranvier after passive
propagation of the pulse through myelinated part of axon, would require extra effort,
and it is reasonable to expect that features underlying pulse creation and
amplification should be emphasized. Expression in these regions of the
dense membrane undercoating might then suggest that structures adjacent to the
membrane inner surface take part in creation and amplification of the pulse, possibly, as hypothesized
above, by a forceful circumferential contraction of a protein
network included in the undercoating.
Another well-known feature of the nodes of Ranvier, high density of voltage-gated sodium
channels, can be justified within the model by necessity to rapidly
change transmembrane voltage by the sodium current to quickly open voltage-gated calcium
channels, by an increase in membrane depolarization at the next Ranvier
node due to augmented axoplasmic currents, and possibly by an increase in
number of sodium channels mechanically coupled to interior
cytoskeleton structures, thus improving mechanosensitivity of the membrane to 
axoplasmic disturbances.

In the model being presented axon acts as a sensor of mechanical
disturbances in the axoplasm, transducing them into pressure-generating motion by
utilizing mechanosensitivity of the membrane ion channels and ionic
electrochemical gradients across the membrane. Considering preponderance
of mechanical stimuli in nature, an appealing scheme of how cells could
have evolved to transmit signals by pressure waves can be constructed.

Let's suppose that a cell has developed an ability to rapidly withdraw a
part of its membrane to which an external negative pressure and/or stretch
was applied. This reaction could be aimed at detachment from adhesion
surfaces, or simply at preservation of shape of a freely moving
cell in a dynamic mechanical environment. If such cell happens to have
cylindrical shape, and an external stimulus causes rapid membrane
withdrawal at a cell end, a wave of increased cytoplasmic pressure will be
created and will propagate along the axis of the cell toward its other
end. Since passive membrane movement is determined by pressure difference
across it, circular membrane segments subjected to increased intracellular pressure in
the pulse will move as if negative pressure was applied at the
extracellular side; membrane stretch caused by the motion of the viscous
cytoplasm in the pulse should also be indistinguishable from a membrane stretch exerted 
from outside the cell. If these stimuli are sufficient to
elicit forceful inward retraction of the circular membrane segments, 
new cytoplasmic pressure waves will
be created, which, if timed properly, will amplify the original
hydraulic pulse carrying the signal of stimulation at one cell
end to the other.
\section{Passive propagation of hydraulic pulses through axons}
For the case of small-amplitude harmonic waves in incompressible liquid
enclosed in a thin-walled elastic tube, with the wave wavelength
large compared to the radius of the tube, and the case of ``large'' liquid
viscosity (defined below), speed of propagation of the waves
$v_{inc}$ and decay length $L_{inc}$ (length over which amplitude of the
waves decreases $e$-fold) are \cite{morg54}
\begin{equation}
v_{inc}=\left(\frac{Eh}{2\rho R}\right)^\frac{1}{2}R\left(\frac{\omega \rho}{\mu}\right)^\frac{1}{2}
\frac{1}{(5-4\nu)^\frac{1}{2}},
\end{equation}
\begin{equation}
L_{inc}=\frac{1}{\omega}\left(\frac{Eh}{2\rho R}\right)^\frac{1}{2}R\left(\frac{\omega \rho}{\mu}\right)^\frac{1}{2}
\frac{1}{(5-4\nu)^\frac{1}{2}},
\end{equation}
where $E$ is the Young's modulus of the tube wall material, $h$ is the thickness of
the tube wall, $R$ is the radius of the tube, $\rho$ is the liquid density, $\mu$
is the liquid viscosity, $\omega$ is the wave frequency, $\nu$ is the
Poisson's ratio of the tube wall material.

The condition of ``large'' viscosity (or small tube radius, or low frequency) 
is $R(\omega\rho/\mu)^{1/2} \ll 1$. That this
condition holds well for axons of diameters at least up to $\sim$ 40 $\mu$m can be
seen from the following estimate. Measurements of macroscopic cytoplasm
viscosity with $\sim$ 1 $\mu$m objects freely diffusing or moving under $\sim$ 1 pN forces
yield $\mu \sim 0.2$ Pa~s \cite{alex91,mars01}, while application of larger $\sim$ 1 nN forces
to objects of similar sizes yields $\mu \sim210$ Pa~s \cite{baus99}. 
Duration of the pressure
pulse should be close to that of the contraction that created and amplified
it, which in turn should be determined by duration of influx of
calcium ions, rate of dissipation of free intracellular calcium and time
properties of the presumed process of contraction. Lacking detailed
knowledge of the last two processes, it is here simply assumed that
duration of the pressure pulse is equal to the AP duration, e.g. $\sim$ 0.34 ms for
large-diameter cat myelinated axons at 37.1 $^\circ$C \cite{paint66}, with corresponding
$\omega\approx2\pi/(2\cdot0.34$ ms$)\approx 9200$ rad/s. Taking $R = 20$ $\mu$m,
$\rho$ = 1000 kg/m$^3$, $\mu$ = 0.2 \mbox{Pa s}, $\omega$ = 9200 rad/s, yields
$R(\omega\rho/\mu)^{1/2}\approx0.14$, which should be smaller enough than 1 for the
large viscosity limit to hold. Another condition used to derive (1) -- (2), of
large wavelength $\lambda$ of the pressure waves compared to the radius of the tube, or
$\lambda = 2\pi v/\omega\gg R$, as will be seen later, also holds well for axons.

For the case of small-amplitude harmonic waves in a compressible liquid enclosed 
in a rigid tube, and the above conditions of large viscosity 
($R(\omega\rho/\mu)^{1/2} \ll 1$) and large wavelength ($\lambda \gg R$) \cite{rayl45}
\begin{equation}
v_{comp}=v_{sound} R\left(\frac{\omega \rho}{\mu}\right)^\frac{1}{2}
\frac{1}{2},
\end{equation}
\begin{equation}
L_{comp}=\frac{1}{\omega} v_{sound} R\left(\frac{\omega \rho}{\mu}\right)^\frac{1}{2}
\frac{1}{2},
\end{equation}
where $v_{sound}$ is the speed of sound in the liquid in the absence of viscosity.

It should be noted that in the absence of viscosity, speed of propagation
of small-amplitude harmonic pressure waves in a compressible liquid
enclosed in a thin-walled elastic tube is given by (e.g. \cite{chev93})
\begin{equation}
v_{\mu=0}=\left(\frac{1}{\rho(k+\frac{2R}{Eh})}\right)^\frac{1}{2},
\end{equation}
where $k$ is the intrinsic bulk compressibility of the liquid, and $k+2R/Eh$ is
the compressibility of the liquid in the tube stemming from both the
intrinsic compressibility of the liquid and distensibility of the walls. In
the limit of rigid tube walls so that their distensibility can be ignored, i.e.
when $k\gg2R/Eh$, (5) reduces to $v_{\mu=0}=(1/\rho k)^{1/2}$, which is equal to the speed
of sound in the liquid $v_{sound}$ (e.g. \cite{chev93}) and the first factor in (3) and 
the second in (4). In the opposing limit
of  $k\ll2R/Eh$, i.e. when internal compressibility of the liquid can be
ignored, (5) simplifies to $v_{\mu=0}=(Eh/2R\rho)^{1/2}$, which is known as the Moens-Korteweg equation 
\cite{hard62} and 
the first factor in (1)
and the second in (2). It now can be seen that outcome of introduction of 
large liquid viscosity in the inviscid case is
independent from whether compressibility of the liquid in the tube arises
from distensibility of the walls, or from intrinsic compressibility of the liquid, 
except for an additional factor $2/(5-4\nu)^{1/2}$ in (1) and (2),
which takes into account movement of the tube wall in the direction of the
tube axis by the viscous liquid \cite{morg54}. In principle it is a straightforward,
although cumbersome, task to modify equation of conservation of mass
(3) in \cite{morg54} to include intrinsic compressibility of the liquid in the
approximate form $d\rho=k\rho dp$, repeat derivations and obtain expressions
for the wave speed and decay length for the case of a compressible viscous liquid
in an elastic tube. Here, however, it is noted that the factor
$2/(5-4\nu)^{1/2}$ varies from 0.89 to 1.15 for physical values of
Poisson's ratio $0<\nu< 0.5$, and is close enough to 1 to be neglected in
the present analysis. Therefore, neglecting axial motion of the tube
wall, (1) -- (5) can be combined to obtain for the case of
small-amplitude harmonic waves in a compressible liquid enclosed in a thin-walled elastic
tube, with the conditions of large liquid viscosity and large wavelength:
\begin{equation}
v=\left(\frac{1}{\rho(k+\frac{2R}{Eh})}\right)^\frac{1}{2}R\left(\frac{\omega \rho}{\mu}\right)^\frac{1}{2}\frac{1}{2},
\end{equation}
\begin{equation}
L=\frac{1}{\omega}\left(\frac{1}{\rho(k+\frac{2R}{Eh})}\right)^\frac{1}{2}R\left(\frac{\omega \rho}{\mu}\right)^\frac{1}{2}\frac{1}{2}.
\end{equation}

In general, when velocity of harmonic waves (phase velocity) $v$ depends on
frequency of the waves $\omega$, velocity of propagation of a wave packet
is equal to the group velocity $v_{gr}$, given by (e.g. \cite{chev93})
\begin{equation}
v_{gr}=v+\omega\frac{dv}{d\omega}.
\end{equation}
Application of (8) to (6) yields group velocity $v_{gr}=1.5v$.  

Derivations in \cite{morg54} assumed isotropy of the tube wall material, while
structure of lipid membranes suggests high anisotropy of their elastic
properties in the transverse direction compared to the in-plane
directions. By following the derivations, however, one can ascertain that
terms corresponding to the transverse elasticity of the wall material have to be
neglected. Then Young's modulus $E$ and Poisson's ratio $\nu$ in the
formulas above represent membrane resistance to a tangential to the membrane stretch and
shrinkage of perpendicular to stretch in-plane dimensions respectively,
under assumption of in-plane isotropy of membrane elastic properties.
Then, neglecting transverse stresses on the membrane, Hook's law yields (e.g. \cite{timo34}):
$e_1=(\sigma_1-\nu\sigma_2)/E$, $e_2=(\sigma_2-\nu \sigma_1)/E$, where $e_1$, $e_2$ --
relative orthogonal elongation of membrane subjected to orthogonal
tangential stresses $\sigma_1$, $\sigma_2$. From the definition of elastic area
expansion modulus $K$ (e.g. \cite{thom97}):  
$(\tau_1+\tau_2)/2=KdA/A$, where $\tau_1$ and $\tau_2$
are orthogonal tensions in the membrane surface, $dA/A$ -- fractional
area change. Taking into account that $\tau_1=\sigma_1h$, $\tau_2=\sigma_2h$, and
for small $e_1$, $e_2$: $dA/A=e_1+e_2$, one can obtain:
\begin{equation}
Eh=2K(1-\nu).
\end{equation}

\section{Numerical estimates and comparison of model predictions to experiment}
Inspection of formulas (6) -- (8) in relation to propagation of pressure
pulses through axons leads to the following statements: velocity of
propagation $v$ for a given $\omega$, $\mu$, $k$, $R$ increases with increased
membrane rigidity $Eh$; for a given $\omega$, $\mu$, $k$, $Eh$, dependence on the axon radius
is intermediate between $R^{1/2}$ and $R$ and specifically depends on the product $Ehk$;
for a limiting case of a soft membrane ($Eh\ll2R/k$), $v\sim R^{1/2}$; for the opposing
case of a rigid membrane ($Eh\gg2R/k$), $v\sim R$; for an axon of fixed $R$, $Eh$ and $k$:
$v\sim(\omega/\mu)^{1/2}$; decay length $L$ behaves similarly to $v$, except for
the $L\sim\omega^{-1/2}$ dependence.

As has been mentioned earlier, the local circuit theory is unable to
explain increased duration of the AP in the smaller diameter fibers of a
related class \cite{hill77}. Since in the model being presented membrane channels are
activated not only by voltage across membrane (as in the local circuit
theory), but also by the traveling pressure pulse, hydraulic pulses of
longer duration should increase duration of the AP due to longer
mechanostimulation of the membrane channels. Since lower frequency components
of a pressure pulse decay less with distance (from (7)), increased damping in
fibers of smaller diameters (again from (7)) in general means that
propagating pressure pulses in these fibers will have lower frequency and longer duration,
and, as mentioned above, longer duration of the AP.

From the above considerations
it also follows that duration of a ``membrane'' AP in this model is predicted to
be independent from the fiber diameter, and equal to
values obtained with the local circuit theory, owing to absence of a
propagating pressure pulse in this case. Duration of a propagated AP,
however, is predicted to be longer than that for a membrane AP,
and is predicted to increase with decreasing fiber diameter reflecting
increased duration of the pressure pulse. The idea that the AP duration $\Delta T_{AP}$
adjusts to the duration of the pressure pulse $\Delta T_{pr}$ reinforces the
approximate relation  $\Delta T_{AP}\approx\Delta T_{pr}$, assumed earlier from converse
arguments that duration of the pressure pulse should be close to that of
influx of calcium ions.
\subsection{Myelinated axons}
As follows from (6) -- (8), the larger the stiffness of the axolemma,
the larger the speed of propagation of the pressure waves and the lesser their decay
with distance. From this perspective, myelinization of axons by Schwann or
oligodendroglial cells tightly wrapping around axon in a spiral manner is
aimed at increase of rigidity of the axon membrane. Successive layers of
the same myelin cell are known to firmly adhere together by proteins from
the cell surface glycoprotein family, which are in fact main protein
constituents of myelin \cite{shep94}. This adherence should be effective
against unwinding of the spiral when subjected to increased
pressure in the hydraulic pulse, and should transfer the distending load to
lipids and proteins of all myelin layers. In order to be
able to neglect distensibility of the myelin sheath compared to
intrinsic compressibility of the axoplasm, ``effective'' Young's modulus $E$ of the
myelin should satisfy stemming from (6) approximate condition
$E>2R/(hk)$. Taking $k$ as for saline water at 37 $^\circ$C, 
$k=(\rho v_{sound}^2)^{-1}\approx4.04\cdot10^{-10}$ Pa$^{-1}$ \cite{lide02}, 
and $R\approx h$ leads to the condition $E>5\cdot10^9$ Pa. Area
expansion modulus $K$ for plasma membranes is on the order of 0.4 N/m with
membrane thickness of $h\approx6$ nm \cite{thom97}, which corresponds 
to $Eh\approx0.6$ N/m (from (9), $\nu$
taken as 1/4) and ``effective'' Young's modulus of $E=Eh/h\approx10^8$ Pa.
Thus, if myelin sheath is as elastic as plasma membrane, it cannot be
considered indistensible compared to the intrinsic compressibility of
axoplasm.  However, considering that lipid composition of myelin differs
markedly from that of plasma membranes, in particular in that it contains
significantly more cholesterol known to increase membrane rigidity \cite{smit96,thom97}, that
the protein-rich major dense lines might also significantly contribute to
myelin rigidity, and that in the model being presented increased
myelin stiffness is beneficial for propagation of the pressure pulses 
and thus should be sought after
by the cellular mechanisms, it is here assumed that circumferential
distensibility of myelin sheath does not appreciably effect velocity of
propagation of axoplasmic pressure waves. Then, also neglecting reduction of
the propagation velocity in the nodes of Ranvier, the quotient $2R/(Eh)$ in (6) and (7) can be
disregarded for myelinated axons.

Substituting in (6) compressibility of saline water $k=4.04\cdot10^{-10}$ Pa$^{-1}$, viscosity
$\mu=0.2$ Pa~s,  frequency $\omega=2\pi/(2\cdot0.6$ ms$)\approx5200$ rad/s, 
axon radius $R=1$ $\mu$m leads to the pressure pulse velocity $v_{gr} = 6$ m/s, which, given
the approximate nature of the estimate, is in a good agreement with the
experimental value of 10 m/s for the AP conduction velocity in cat
myelinated axons of 1 $\mu$m radius at 38 $^\circ$C \cite{hill77}; 
the AP duration of 0.6 ms for the
mentioned fibers was measured at 37.1 $^\circ$C in \cite{paint66}. 
Use of $\mu=0.2$ Pa~s in the
estimate can be justified by considering that it should represent viscosity
arising from small-amplitude cytoskeletal deformations \cite{alex91,mars01} that are likely
to accompany propagation of a small-amplitude pressure pulse, while larger
values of $\mu=210$ Pa~s were obtained with forceful displacement of
cytoplasmic objects that possibly disrupted the cytoskeleton or distorted it
considerably \cite{baus99}; still, given the large scatter of values for $\mu$ 
available in the literature, uncertainty in $\mu$ should represent a major contribution
to overall error of the estimate.

From (7), $e$-fold amplitude decay length for the considered case is then $L\approx1.2$ mm, and
wavelength $\lambda=2\pi v/\omega\approx7.3$ mm.
\subsection{Unmyelinated axons}
For unmyelinated axons of radius $R=0.65$ $\mu$m, assuming membrane $Eh=0.6$ N/m
as before, and taking 
$\mu=0.2$ Pa~s, $k=4.04\cdot10^{-10}$ Pa$^{-1}$, $\omega=5200$ rad/s as 
for the myelinated axons, (6) and (8) yield 
the pressure pulse velocity $v_{gr}=0.053$ m/s, which is $\sim$40 times lower compared to
measurements of 2.3 m/s for the AP conduction velocity in cat unmyelinated nerve fibers 
of 0.65 $\mu$m radius at 38 $^\circ$C \cite{hill77}, notwithstanding
the low value for viscosity used. Taking into account agreement obtained with the same values
of $\mu$, $k$, $\omega$ for 
myelinated axons, and that the main difference in mechanical properties between
myelinated and unmyelinated axons should be in the axolemmal resistance $Eh$ to 
increases in axon diameter, one can conclude that in order for the model to predict correct AP
conduction velocity for the unmyelinated axons, ``effective'' membrane $Eh$ has to be 
$\sim1600$ times larger than the value
of $0.6$ N/m used. This latter value represents resistance of
lipid plasma membranes to area expansion. Additional resistance to 
increases in diameter of unmyelinated axons can come from deformation of adjacent glial cells; 
from deformation of the cytoskeleton tethering the membrane; 
from passive resistance of the membrane skeletal protein
network to the radial deformation and from active contraction of the network during 
amplification of the hydraulic pulse. Since axial membrane elasticity does
not appreciably influence velocity and decay length of the pressure waves, 
the mentioned above factors might selectively increase membrane resistance to the
circumferential expansion, leaving axial rigidity at a lower value. 
From this perspective, the model predicts ``effective''  
area expansion modulus for the deformation described $\sim$ 3 orders of magnitude
larger than for lipid membranes.

\subsection{Velocity dependence on axon diameter and temperature}
From (6) -- (8) it follows that for unmyelinated fibers of small
diameters ($2R\ll Ehk$), $v_{gr}\sim R$,
while in the opposing limit of large diameters ($2R\gg Ehk$),
$v_{gr}\sim R^{1/2}$. 
In unmyelinated C-fibers (diameters 0.4 -- 1.2 $\mu$m) of cat saphenous nerve AP conduction 
velocity $v_{AP}$ is proportional to $R$, in unmyelinated fibers
(1.6 -- 20 $\mu$m) of locust and cockroach $v_{AP}\sim R^{0.7}$ -- $R^{0.8}$, while in unmyelinated
fibers (2 -- 520 $\mu$m) of cephaloid molluscs $v_{AP}\sim R^{0.57}$ \cite{hill77}. 
Thus, assuming that
$Ehk$ is on the order of 10 $\mu$m and is not significantly different 
for the mentioned fibers, predicted
gradual change in dependence from $R$ to $R^{0.5}$ with increasing unmyelinated
fiber diameter is very well matched by experiment.

For myelinated fibers, assuming $Eh\gg2R/k$, (6) -- (8) give $v_{gr}\sim R$ dependence, 
again closely matched by experimentally determined $v_{AP}\sim R$ \cite{hill77}. 
In the considered above 
limit of small unmyelinated fibers, condition $2R\ll Ehk$ is the same as for myelinated fibers,
$Eh\gg2R/k$, which indicates that for these fibers myelinization would not
increase the speed of propagation or decay length. Very small diameter fibers
are indeed usually unmyelinated \cite{rush51}. 

The analyses above implicitly assumed
that specific mechanical properties of axons, $k$, $\mu$, $Eh$, 
are constant across fibers of varying diameters, or
that their variation is small enough not to appreciably change
obtained $R$ dependences. The same condition was assumed for the central
frequency $\omega$ of the pressure wave packet. In cat saphenous myelinated 
fibers at 37.1 $^\circ$C, AP duration $\Delta T_{AP}$ changes from 0.6 ms to 0.34 ms with 8-fold
increase in diameter \cite{paint66}. Corresponding change in $\omega=2\pi/(2\cdot\Delta T_{AP})$, 
from $5.2\cdot10^3$ to $9.2\cdot10^3$ rad/s, can be shown to introduce
not more than an additional 0.14 power of $R$ in the predicted dependences. 
A larger variation of $\omega$ with $R$, however, would have a 
greater impact on the predictions.

Assuming that $Eh$, $k$ and $R$ do not significantly depend on temperature $T$,
and that changes in $\omega$ and $\mu$ with $T$ are independent from 
degree of fiber myelinization, (6) -- (8) predict that myelinated and
unmyelinated fibers should be similarly
affected by a temperature change, just as observed experimentally \cite{hill77};
the predicted dependence is $v_{AP}\sim(\omega(T)/\mu(T))^{1/2}$.
Experimental data on $\mu(T)$ is not readily
available, however, for most liquids viscosity increases with decreasing
temperature \cite{poli01}. An increase in viscosity means (see (7)) 
augmented damping of pressure waves of all
frequencies, and, just as in the case of fibers of smaller diameters
considered earlier, duration of the AP can be argued to increase. AP
duration indeed increases with decreasing temperature, e.g. with Q$_{10}$ of 3.4 at
37 $^\circ$C for cat vagus myelinated fibers \cite{paint66}, but this large variation
could mainly stem from change in rate of activation of voltage-gated 
membrane channels \cite{hill77}.
According to $v_{gr}\sim\omega^{1/2}$ and the assumed relation 
$\omega=2\pi/(2\cdot\Delta T_{AP})$,
Q$_{10}$ of 3.4 for the AP duration $\Delta T_{AP}$ implies that Q$_{10}$ for
the pressure pulse propagation velocity is predicted to be $3.4^{1/2}\approx1.8$ 
without even considering $\mu(T)$ dependence. Experimental value of $Q_{10}=1.6$ 
for the AP conduction velocity in the cat vagus 
myelinated fibers at 37 $^\circ$C \cite{paint65} thus might indicate that viscosity of
the axoplasm does not change appreciably in the 27 -- 37 $^\circ$C range.

It is also interesting to note that cooling-related removal of AP conduction block 
in damaged (e.g. by multiple sclerosis) myelinated axons \cite{fran99} can be 
naturally explained within the model by increase in the decay length $L$ with the
decrease of $\omega\sim(\Delta T_{AP})^{-1}$ at lower temperatures.
\subsection{Depletion of sodium and calcium concentrations}
Depletion of sodium concentration around a squid giant axon leads
to wider and slower propagating APs \cite{hodg49}, in close
agreement with the Hodgkin-Huxley equations \cite{hill92}. However,
longer duration of the AP in the presented model also decreases velocity of
AP propagation, through relation $v_{AP}\sim(\Delta T_{AP})^{-1/2}$; qualitative
analysis of $v_{AP}$ and $\Delta T_{AP}$ from Fig. 4 in
\cite{hodg49} does not exclude validity of this relation. 

Conduction in axons in low external calcium is poor or blocked at turnon of cathodal current \cite{hill77}. These axons, however, do 
respond better to anodal break excitation \cite{hill77}, which does provide higher gradient for
Ca$^{2+}$ to enter the axon. Of interest is whether axons propagate APs in extremely depleted external calcium. 
If they do, interesting is to know whether conduction velocity increases or decreases, the latter being predicted
by the model, if conduction can switch to lower velocity local circuit mechanism.
In the local circuit theory, conduction velocity with depleted external calcium should, in the first
approximation, increase: decrease 
in free extracellular calcium significantly
decreases thresholds of activation of both sodium and potassium channels \cite{hill92}; 
free intracellular calcium can activate polarizing calcium-dependent 
potassium currents \cite{hill92}. Then, observed increase in the velocity would favor
the local circuit theory, but, strictly speaking, would not disprove
the model, since it is possible that
the velocity given by the local circuit theory became larger 
than that predicted by the model after modification of channel gating, and the local
circuit theory mechanism took over the propagation from the pressure wave. Data on 
the AP conduction velocities in axons with manipulated calcium concentrations
is not readily available, since most experiments use 
voltage-clamp methods under these conditions; detailed analysis of these phenomena
will be done in a subsequent work.

\section{Discussion}
It is very remarkable that transmission of hydraulic pulses through
axoplasm, a process so disparate from passive spread of ionic
currents, reasonably well reproduces all major features of AP
conduction: changes in the velocity with membrane rigidity, with axon
diameter, with the temperature and with AP duration; changes in the AP duration with fiber
diameter. Accurate predictions of absolute AP conduction velocities by the model are
limited by the knowledge of relevant to
propagation of the pressure pulses cytoplasm viscosity, resistance of myelinated
and unmyelinated axons to increases in diameter,
cytoplasm compressibility, and by approximate nature of the
estimate of duration of the hydraulic pulse. Still, estimates of
the velocity for myelinated fibers, with an assumption of indistensibility of the myelin sheath,
give values close to those observed experimentally; for unmyelinated 
fibers, although the velocity is underpredicted, the situation might change
as the appropriate experimental data on membrane distensibility becomes available.

It also has been noted (Prof. M. Kardar, 2003, private communication) that 
the lipid membrane can be more realistically
modeled as 2-dimensional liquid rather than elastic solid, with wave propagation velocity shaped by
bending membrane rigidity rather than area expansion modulus. With the different treatment, which 
will be attempted in the future,
predicted by the model AP velocity in unmyelinated axons might change dramatically.

On the other hand it is noteworthy that all fundamental ingredients
necessary for realization of the described model are in place:  
mechanosensitivity of many membrane ion channels (e.g. \cite{gu01}), 
low free calcium concentration in a resting cell, ubiquity of the calcium
impulse and its trait to induce motion in filaments \cite{shep94}, and high filament
content of the cell interior \cite{shep94}. Whether or not these ingredients are present
to a degree sufficient for the active propagation of the hydraulic pulses,
and whether any cells in fact use the described mechanism for AP
conduction, should be best settled by experiment.

A small axon swelling followed by a smaller shrinkage, both roughly synchronous with
propagating APs, was demonstrated in crab and
squid giant axons \cite {iwas80}.
Beyond increased pressure inside axon, the swelling
can be caused by variation of the voltage across membrane through the
ensuing change in difference in surface tensions between the two membrane
interfaces \cite{zhan01}. Axial motion of an axoplasmic object, such
as a vesicle, observed simultaneously with the AP, should be less influenced
by the voltage-dependent membrane movement and more sensitive to
axoplasm disturbances induced by the hydraulic pulse. The model predicts
that a swelling should
precede the voltage spike, and, 
in fact, figures in \cite{iwas80} indicate that the swelling 
by several milliseconds does precede the voltage spike, but the authors'
conclusion was that cause of that might be in electronics delays;
careful analysis of the delays in an experiment similar to that in \cite{iwas80} could
provide another test of the model. 

Knockout of a neurofilament subunit gene has been
reported to $\sim$ twice decrease conduction velocity in mice large myelinated axons,
with axon diameter, AP amplitude, duration and shape preserved \cite{kriz00} .
Neurofilament proteins are the most abundant cytoskeletal element in large myelinated axons \cite{kriz00}.
These data are favorable for the presented model, with the changes in conduction velocity explained by modification 
of cytoskeletal and membrane skeletal protein mechanics. The local circuit theory predicts
that conduction velocity is independent from mechanical properties of axons. Somewhat contrary to that,
it is known that cell pressurization does block AP conduction in at least in some axons \cite{hill77}.

Other tests can come from measurements
of velocity of propagation of pressure pulses through axons; from precise
measurements of mechanical properties of neurons and other excitable cells; from
experiments on mechanical stimulation of AP in excitable cells;
from measurements of the conduction velocity with
depletion of free intracellular calcium, or with disruption of the membrane skeletal proteins;
or from a setup aimed at direct detection of pressure waves accompanying APs.

The following circumstances can be recognized to indirectly support the
model: in many non-neuronal excitable cells AP seems to be connected to
mechanical motion \cite{shep94}; neurons often have convex shape as if pressurized
from inside; myelinization appears to be the only feasible way to
substantially increase mechanical resistance of axolemma, while
resistance of the membrane to leak currents could in principle be
increased by other means, e.g. decrease in number of leak channels;
many freely moving cells have an ability to control membrane movement by
forces of microfilaments located just beneath the plasma membrane \cite{shep94}; the
same ability is attributed to neurons, where the forces are believed to
arise from interactions of actin, fodrin and other elements \cite{shep94}. Finally,
the main components of the model, propagating hydraulic pulses
and radially contracting tubes generating a directed axial force on the
tube contents, are exploited in cardiovascular, intestinal and other
tubular systems in many organisms, which might even suggest an
evolutionary link between the systems and excitable cells.

Central role of hydraulic pulses and pressure-generating processes in 
the model implies that pressure-related
phenomena should play important roles in the function of those neurons and
excitable cells, that would realize the model mechanisms: 
synaptic vesicle release in neurons might be aided by 
impact of the AP-associated hydraulic pulse on the vesicles or
by contraction of the bell-shaped membrane skeletal protein network
of the synapse, creating a directed
push on the cytoplasm and the vesicles; depolarization of the axon hillock 
to the excitation level might be accomplished not
only by passive spread of ionic currents from postsynaptic sites, but also 
by mechanoactivation of the membrane ion channels by 
cytoplasm displacement or pressurization; therefore, synaptic integration might involve relay of
displacement or pressurization of cytoplasm through the dendritic tree to the axon hillock, and
learning and memory in neurons might
be partially reflected in strength and duration of Ca$^{2+}$-activated 
contraction of postsynaptic membrane skeletal
proteins, or in another pressure-generating process; 
AP generated in a neuron might influence processes in the cell body, dendrites and 
contacting synapses by mechanical disturbances introduced by the back-propagating pressure pulse;
AP-associated hydraulic pulses might participate in transport of substances in excitable cells. 
Other outcomes of the model are the hypotheses on the evolution of the signal transmission
in cells, on reasons for mechanosensitivity of many unrelated types of ion 
channels, and on the physiological role of axonal calcium influx.

In the case the presented model is not experimentally confirmed, calculations 
in the paper show that velocity of propagation of $\sim$ 1 kHz pressure waves 
through axons can be close to the AP conduction velocities, and hence
pressure waves generated by voltage-induced membrane movement during AP
conduction \cite{zhan01} might accumulate into a larger shock-like wave.
\begin{acknowledgements}The author gratefully thanks his father Prof. Michael A. Rvachov for 
reminding of the idea, Prof. William Bertozzi for unwavering help and 
support, and 
Prof. Guosong Liu for a constructive discussion.
While working on the project the author was supported by US DOE grant DE-FC02-94ER40818.
\end{acknowledgements}

\bibliography{proj}

\end{document}